\documentclass[aps, floatfix,prl,twocolumn,showpacs,superscriptaddress]{revtex4-2}
\usepackage{graphicx}
\usepackage{dcolumn}
\usepackage{bm}

\usepackage{soul}
\usepackage{color}
\usepackage{url}
\usepackage{hyperref}
\usepackage{blindtext} 
\usepackage[normalem]{ulem}
\usepackage{orcidlink}

\def\tam{\textsuperscript{180m}Ta}
\def\tag{\textsuperscript{180}Ta}
\def\taH{\textsuperscript{182}Ta}

\def\hffive{\textsuperscript{175}Hf}

\def\MJ{{\sc Majorana}} 
\def\DEM{{\sc Demonstrator}} 
\def\MJD{{\sc Majorana Demonstrator}}

\begin{document}

\title{Constraints on the decay of \textsuperscript{180m}Ta}

\newcommand{\ITEP}{National Research Center ``Kurchatov Institute'', Kurchatov Complex of Theoretical and Experimental Physics, Moscow, 117218 Russia}
\newcommand{\JINR}{Joint Institute for Nuclear Research, Dubna, 141980 Russia}
\newcommand{\lbnl}{Nuclear Science Division, Lawrence Berkeley National Laboratory, Berkeley, CA 94720, USA}
\newcommand{\lbnle}{Engineering Division, Lawrence Berkeley National Laboratory, Berkeley, CA 94720, USA}
\newcommand{\lanl}{Los Alamos National Laboratory, Los Alamos, NM 87545, USA}
\newcommand{\queens}{Department of Physics, Engineering Physics and Astronomy, Queen's University, Kingston, ON K7L 3N6, Canada}
\newcommand{\uw}{Center for Experimental Nuclear Physics and Astrophysics, and Department of Physics, University of Washington, Seattle, WA 98195, USA}
\newcommand{\unc}{Department of Physics and Astronomy, University of North Carolina, Chapel Hill, NC 27514, USA}
\newcommand{\duke}{Department of Physics, Duke University, Durham, NC 27708, USA}
\newcommand{\ncsu}{Department of Physics, North Carolina State University, Raleigh, NC 27695, USA}
\newcommand{\ornl}{Oak Ridge National Laboratory, Oak Ridge, TN 37830, USA}
\newcommand{\ou}{Research Center for Nuclear Physics, Osaka University, Ibaraki, Osaka 567-0047, Japan}
\newcommand{\pnnl}{Pacific Northwest National Laboratory, Richland, WA 99354, USA}
\newcommand{\ttu}{Tennessee Tech University, Cookeville, TN 38505, USA}
\newcommand{\sdsmt}{South Dakota Mines, Rapid City, SD 57701, USA}
\newcommand{\usc}{Department of Physics and Astronomy, University of South Carolina, Columbia, SC 29208, USA}
\newcommand{\usd}{Department of Physics, University of South Dakota, Vermillion, SD 57069, USA}
\newcommand{\ut}{Department of Physics and Astronomy, University of Tennessee, Knoxville, TN 37916, USA}
\newcommand{\tunl}{Triangle Universities Nuclear Laboratory, Durham, NC 27708, USA}
\newcommand{\mpi}{Max-Planck-Institut f\"{u}r Physik, M\"{u}nchen, 80805, Germany}
\newcommand{\tum}{Physik Department and Excellence Cluster Universe, Technische Universit\"{a}t, M\"{u}nchen, 85748 Germany}
\newcommand{\williams}{Physics Department, Williams College, Williamstown, MA 01267, USA}
\newcommand{\ciemat}{Centro de Investigaciones Energ\'{e}ticas, Medioambientales y Tecnol\'{o}gicas, CIEMAT 28040, Madrid, Spain}
\newcommand{\iu}{IU Center for Exploration of Energy and Matter, and Department of Physics, Indiana University, Bloomington, IN 47405, USA}
\newcommand{\Stanford}{Stanford Institute for Theoretical Physics, Stanford University, Stanford, CA 94305, USA}

\author{I.J.~Arnquist}\affiliation{\pnnl}
\author{F.T.~Avignone~III}\affiliation{\usc}\affiliation{\ornl}
\author{A.S.~Barabash\,\orcidlink{0000-0002-5130-0922}}\affiliation{\ITEP}
\author{C.J.~Barton}\affiliation{\usd}
\author{K.H.~Bhimani}\affiliation{\unc}\affiliation{\tunl}
\author{E.~Blalock\,\orcidlink{0000-0001-5311-371X}}\affiliation{\ncsu}\affiliation{\tunl}
\author{B.~Bos}\affiliation{\unc}\affiliation{\tunl}
\author{M.~Busch}\affiliation{\duke}\affiliation{\tunl}
\author{M.~Buuck\,\orcidlink{0000-0001-5751-4326}}\affiliation{\uw}
\author{T.S.~Caldwell}\affiliation{\unc}\affiliation{\tunl}
\author{C.D.~Christofferson}\affiliation{\sdsmt}
\author{P.-H.~Chu\,\orcidlink{0000-0003-1372-2910}}\affiliation{\lanl}
\author{M.L.~Clark}\affiliation{\unc}\affiliation{\tunl}
\author{C.~Cuesta\,\orcidlink{0000-0003-1190-7233}}\affiliation{\ciemat}
\author{J.A.~Detwiler\,\orcidlink{0000-0002-9050-4610}}\affiliation{\uw}
\author{Yu.~Efremenko}\affiliation{\ut}\affiliation{\ornl}
\author{H.~Ejiri}\affiliation{\ou}
\author{S.R.~Elliott\,\orcidlink{0000-0001-9361-9870}}\affiliation{\lanl}
\author{G.K.~Giovanetti}\affiliation{\williams}
\author{J.~Goett}\affiliation{\lanl}
\author{M.P.~Green\,\orcidlink{0000-0002-1958-8030}}\affiliation{\ncsu}\affiliation{\tunl}\affiliation{\ornl}
\author{J.~Gruszko\,\orcidlink{0000-0002-3777-2237}}\affiliation{\unc}\affiliation{\tunl}
\author{I.S.~Guinn\,\orcidlink{0000-0002-2424-3272}}\affiliation{\unc}\affiliation{\tunl}
\author{V.E.~Guiseppe\,\orcidlink{0000-0002-0078-7101}}\affiliation{\ornl}
\author{C.R.~Haufe}\affiliation{\unc}\affiliation{\tunl}
\author{R.~Henning}\affiliation{\unc}\affiliation{\tunl}
\author{D.~Hervas~Aguilar}\affiliation{\unc}\affiliation{\tunl}
\author{E.W.~Hoppe\,\orcidlink{0000-0002-8171-7323}}\affiliation{\pnnl}
\author{A.~Hostiuc}\affiliation{\uw}
\author{I.~Kim}\affiliation{\lanl} \altaffiliation{Present address: Lawrence Livermore National Laboratory, Livermore, CA 94550, USA}
\author{R.T.~Kouzes}\affiliation{\pnnl}
\author{T.E.~Lannen~V}\affiliation{\usc}
\author{A.~Li\,\orcidlink{0000-0002-4844-9339}}\affiliation{\unc}\affiliation{\tunl}
\author{J.M. L\'opez-Casta\~no}\affiliation{\ornl}
\author{R.~Massarczyk\,\orcidlink{0000-0001-8001-9235}}\email[Corresponding author : ]{massarczyk@lanl.gov}\affiliation{\lanl}
\author{S.J.~Meijer\,\orcidlink{0000-0002-1366-0361}}\affiliation{\lanl}
\author{W.~Meijer\,\orcidlink{0000-0002-7608-2378}}\affiliation{\lanl}
\author{T.K.~Oli\,\orcidlink{0000-0001-8857-3716}}\altaffiliation{Present address: Argonne National Laboratory, Lemont, IL 60439, USA}\affiliation{\usd}
\author{L.S.~Paudel\,\orcidlink{0000-0003-3100-4074}}\affiliation{\usd}
\author{W.~Pettus\,\orcidlink{0000-0003-4947-7400}}\affiliation{\iu}
\author{A.W.P.~Poon\,\orcidlink{0000-0003-2684-6402}}\affiliation{\lbnl}
\author{D.C.~Radford}\affiliation{\ornl}
\author{A.L.~Reine\,\orcidlink{0000-0002-5900-8299}}\affiliation{\unc}\affiliation{\tunl}
\author{K.~Rielage\,\orcidlink{0000-0002-7392-7152}}\affiliation{\lanl}
\author{A.~Rouyer\,\orcidlink{0009-0001-1456-0098}}\affiliation{\williams}
\author{N.W.~Ruof\,\orcidlink{0000-0001-9665-6722}}\affiliation{\uw}	\altaffiliation{Present address: Lawrence Livermore National Laboratory, Livermore, CA 94550, USA}
\author{D.C.~Schaper\,\orcidlink{0000-0002-6219-650X}}\affiliation{\lanl}
\author{S.J.~Schleich\,\orcidlink{0000-0003-1878-9102}}\affiliation{\sdsmt}
\author{T.A.~Smith-Gandy}\affiliation{\williams}
\author{D.~Tedeschi}\affiliation{\usc}
\author{R.L.~Varner\,\orcidlink{0000-0002-0477-7488}}\affiliation{\ornl}
\author{S.~Vasilyev}\affiliation{\JINR}
\author{S.L.~Watkins\,\orcidlink{0000-0003-0649-1923}}\affiliation{\lanl}
\author{J.F.~Wilkerson\,\orcidlink{0000-0002-0342-0217}}\affiliation{\unc}\affiliation{\tunl}\affiliation{\ornl}
\author{C.~Wiseman\,\orcidlink{0000-0002-4232-1326}}\affiliation{\uw}
\author{W.~Xu}\affiliation{\usd}
\author{C.-H.~Yu\,\orcidlink{0000-0002-9849-842X}}\affiliation{\ornl}

\collaboration{{\sc{Majorana}} Collaboration}

\author{D.S.M.~Alves\,\orcidlink{0000-0002-4447-6305}}\affiliation{\lanl}
\author{H.~Ramani}\affiliation{\Stanford}

\noaffiliation

\date{\today}

\begin{abstract}
\tam\ is a rare nuclear isomer whose decay has never been observed. Its remarkably long lifetime surpasses the half-lives of all other known $\beta$ and electron capture decays due to the large K-spin differences and small energy differences between the isomeric and lower energy states. Detecting its decay presents a significant experimental challenge but could shed light on neutrino-induced nucleosynthesis mechanisms, the nature of dark matter and K-spin violation. For this study, we repurposed the \MJD, an experimental search for the neutrinoless double-beta decay of $^{76}$Ge using an array of high-purity germanium detectors, to search for the decay of \tam. More than 17 kilograms, the largest amount of tantalum metal ever used for such a search was installed within the ultra-low background \MJD\ detector array. In this paper we present results from the first year of Ta data taking and provide an updated limit for the \tam\ half-life on the different decay channels. With new limits up to $1.5\times 10^{19}$ years, we improved existing limits by one to two orders of magnitude. This result is the most sensitive search for a single $\beta$ and electron capture decay ever achieved.
\end{abstract}

\maketitle

The \tam\ isomer is unique in two interesting ways: it is the only naturally occurring long-lived isomer, and it is the only known isomer that has not been observed to decay while its ground state has a half-life of only 8.15 hours~\cite{GALLAGHER1962285}. This remarkable property can be attributed to a combination of two factors. The large difference in the K-spin which stands for the projection the spin on the symmetry axis; and the small energy differences requiring an \textsl{E7} or \textsl{M8}-transition between the isomeric and lower lying states~\cite{Walker1999,Walker01,Bissell:2006uy}. The combination of both result in the isomer being trapped in a metastable excited state. Over the last century numerous attempts were made to measure the decay of \tam\ ~\cite{Bauminger:1958,Ryves_1980,Cumming:1985zza,Wakasugi:1994,Hult:2006tp,Lehnert:2016iku, 2305.17238}.

\begin{figure}[htbp]
 \centering
\includegraphics[width = 1.0\columnwidth]{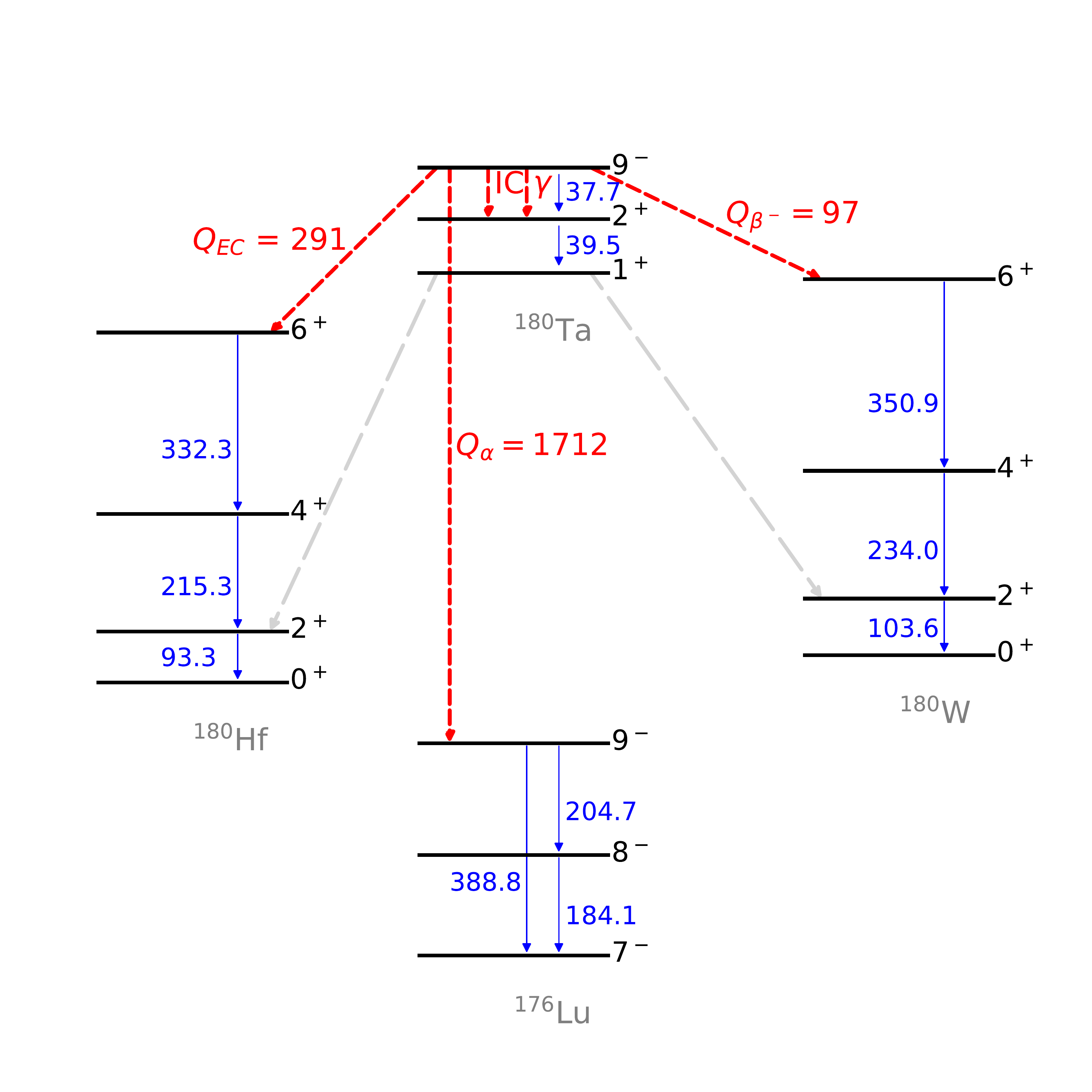}
\caption{Level diagram of the decay modes of \tam\ (red arrows) based on data from Ref.~\cite{NucleDat180}. Certain decay modes can also be observed indirectly when the ground state of \tag\ is populated that then decays further (gray dashed arrows). Emissions of $\gamma$-rays (blue arrows) at characteristic energies can be used to identify the different signatures. The nomenclature follows Eq.~\ref{Eq:1}, all energies are given in\,keV, and all channels except DM de-excitation are depicted.}
\label{fig:levels}
\end{figure}

As shown in Fig.~\ref{fig:levels}, \tam\ has several possible decay modes. These include de-excitation to a lower lying state by $\gamma$-ray emission or internal conversion (IC), electron-capture (EC) decay to $^{180}$Hf ~\cite{Wilkinson:1949}, to $^{180}$W by $\beta^{-}$ decay, and $\alpha$ decay to $^{176}$Lu. As shown in Tab.~\ref{Tab:180Tanumbers} the IC mode is expected to be the fastest decay and the $\gamma$-ray emission is expected to be the slowest~\cite{Auerbach:2017, Ejiri:2017dro}. While often neglected, the possibility of an $\alpha$ decay branch is motivated by a positive Q-value and the observation of $\alpha$ decays with $10^{18}$-year half-lives in neighboring W isotopes~\cite{Cozzini:2004vd}. We follow the common behavior of $\alpha$ decays that similar spin and parity in the daughter are preferred, hence a specific state in $^{176}$Lu is favored, see Fig.~\ref{fig:levels}.
 The total decay width of the \tam\ isomer can expressed as:
\begin{equation}
 \Gamma_{total} = \Gamma_{EC}+\Gamma_{\beta^-}+\Gamma_\gamma+\Gamma_{IC}+\Gamma_{\alpha}+\Gamma_{DM}
 \label{Eq:1}
\end{equation}
Here, $\Gamma_{EC}$ and $\Gamma_{\beta^-}$ are the decay of the isomeric state directly to Hf by EC and $\beta^-$ decay to W. The decay width $\Gamma_\gamma$ and $\Gamma_{IC}$ are isomeric transitions to lower lying states of \tag\ via $\gamma$-ray emission or IC. $\Gamma_{\alpha}$ is the $\alpha$ decay, and $\Gamma_{DM}$ is the decay due to the possible isomeric de-excitation by dark matter (DM)~\cite{Pospelov2020}. Each decay mode can be identified by characteristic $\gamma$-rays, cf. Fig.~\ref{fig:levels}. If no decay is found the total width has to be bigger then the smallest half-life limit whereas $\Gamma=1/T$.

Theoretical techniques~\cite{Ejiri:2017dro, Ejiri:2019ezh} have been proposed for estimating the lifetime of deformed nuclei like \tam. A measurement of the \tam\ decay rate would test the accuracy of these models, particularly the K-selection rule based on the symmetry of the deformation~\cite{Ejiri:2017dro}, under the most extreme conditions. In addition, long-lived isomers can be used to constrain DM models by considering the contributions of DM-induced transitions on the decay rate~\cite{Pospelov2020}. Finally, the measurement of the \tam\ lifetime could help explain the observed abundance of \tag\ and its role within a nucleosynthesis framework~\cite{Mohr:2006wn, Hayakawa10, Baccolo2015, MALATJI2019403}.

Despite being an isotope of interest for almost a century, measuring the decay of the metastable isomer is experimentally challenging. The natural isotopic abundance is very small~\cite{Baccolo2015} and obtaining sufficient quantities of the isotope is difficult. Additionally, the expected energies of the decay emissions are low while the density and atomic number of tantalum metal are high, which makes it challenging to maintain reasonable detection efficiency while increasing the sample mass due to self shielding. Finally, the decay rate is very slow, making standard radioassay techniques insufficient for detection. To overcome these a larger amount material then ever before was installed into the ultra-low background environment of the \MJD.
The purpose of \MJ\ was to demonstrate the feasibility of using high-purity germanium (HPGe) detectors for a ton-scale neutrinoless double-beta decay search in $^{76}$Ge and to explore the low-background experimental techniques required to build such a detector~\cite{Majorana:2013cem,Majorana:2019nbd}. Located at the 4850-ft level of the Sanford Underground Research Facility (SURF)~\cite{Heise_2015}, it consisted of two arrays of HPGe detectors in vacuum cryostats, most of which were enriched in $^{76}$Ge. These were arranged within a passive copper, lead, and polyethylene shield as well as an active muon veto. Data taking with the enriched detectors concluded in 2021~\cite{Majorana:2022udl}. The success of the \DEM\ was enabled by the careful selection and development of ultra low-background components~\cite{Abgrall:2016cct}, the use of low-noise electronics and data acquisition hardware~\cite{Majorana:2021mtz}, and excellent energy resolution achieved through a combination of detector design and novel analysis techniques~\cite{9288848,Majorana:2022vai}. These features also made the \MJ\ an ideal platform for investigating the decay of \tam. In 2022, following the completion of the neutrinoless double-beta decay search and the removal of the enriched detectors for use in LEGEND-200~\cite{LEGEND}, the \DEM\ was repurposed to make this measurement.

To implement tantalum in the existing setup, 99.995\% pure Ta metal disks were purchased from Goodfellow Corp.~\cite{Goodfellow}. Each disc is 2~mm thick with a mass of approximately 181~g. They were brought underground in January 2022, where they underwent a multi-step cleaning process. The discs were scrubbed with Micro-90 to remove oil and manufacturing dirt, then underwent a light chemical etch using 10\% nitric acid, and were finally baked under high vacuum. A total of 17.39~kg of Ta disks were installed within the \DEM, resulting in a total \tam\ mass of 2.045~g, assuming a \tam\ natural abundance of 0.0001176(23)~\cite{PubChem2023}, which is a combined analysis of several previous measurements~\cite{Tant1980,BerglundWieser2011,C6JA00329J, deLaeter:2005kc}.

\begin{figure}[t]
 \centering
 \includegraphics[width=0.4\columnwidth]{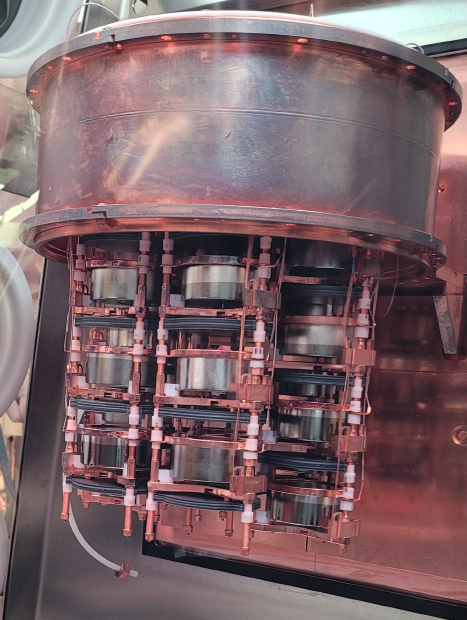}
 \includegraphics[width=0.2\columnwidth]{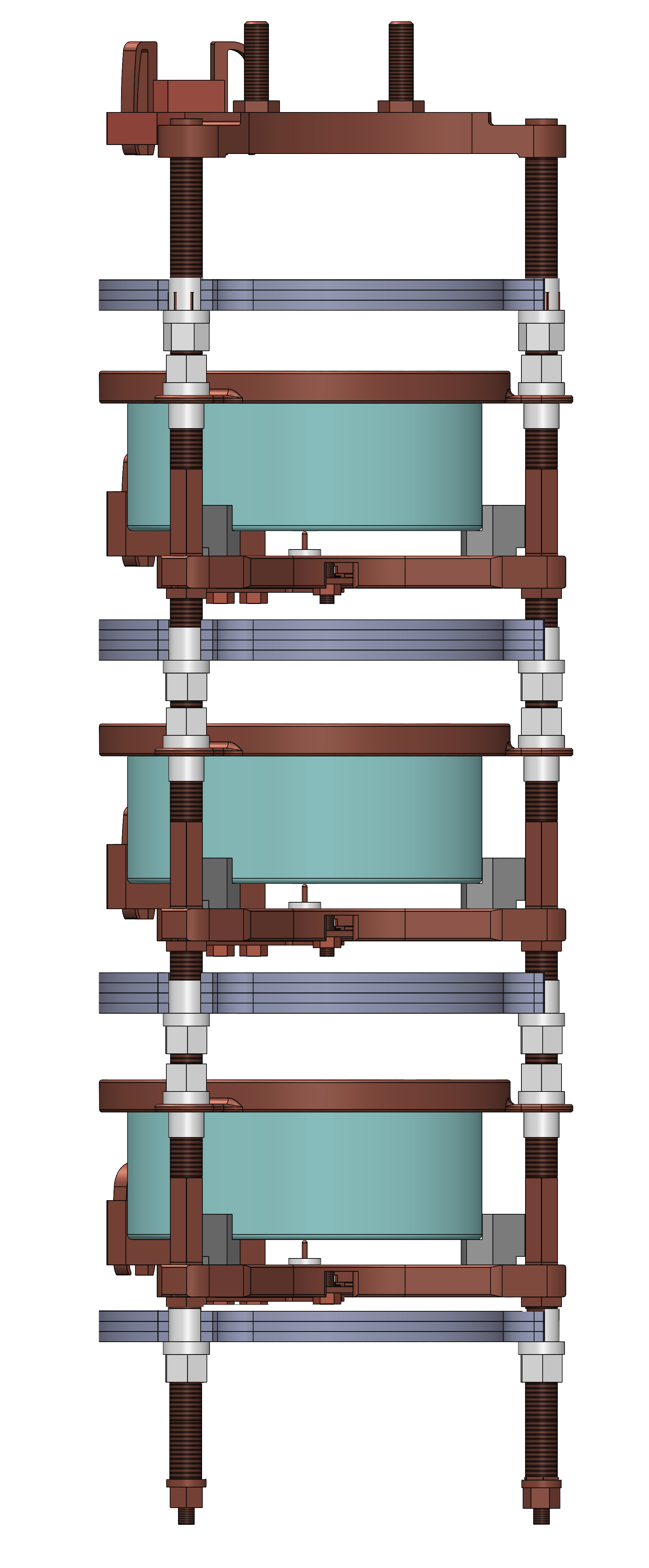}
 \includegraphics[width=0.17\columnwidth]{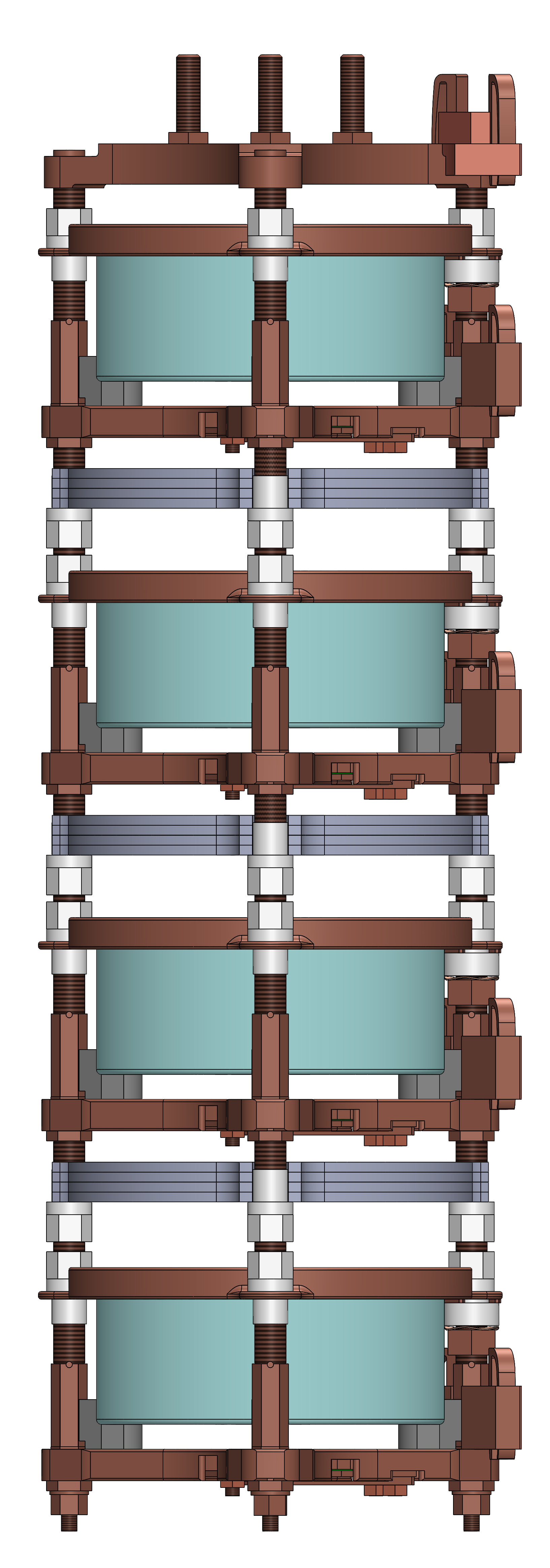}
 \caption{(Left) The detector module during assembly. (Right) Technical drawing of two of the seven installed strings detector arrangement with three and four HPGe detectors (teal) and the tantalum sample disks (gray).}
 \label{FigArrangments}
\end{figure}

To maximize exposure and detection efficiency while preserving the low-background performance of the \DEM, a scheme was developed to re-use previously screened, ultra-high-radiopurity components to hold the Ta samples and interleave them with the 23 remaining natural detectors~\cite{Bege}. A Geant4~\cite{GEANT4:2002zbu} simulation was used to determine the optimal positioning of the samples while respecting the weight and geometry constraints of the \DEM\ cryostat. The final arrangement optimizes the thickness of the Ta samples against efficiency for detecting the low energy $\gamma$-rays of interest. Figure~\ref{FigArrangments} shows the final configuration. Detectors in neighboring strings are offset vertically so that each stack of 3 or 4 Ta discs has a line of sight with at least three HPGe detectors.

This paper presents data collected over 348 days between May 2022 and April 2023. Each of the 23 HPGe detectors in the array is read out independently, in a similar fashion to the \MJD\ experiment. Detector waveforms that exceed approximately 5~keV are digitized with GRETINA digitizers~\cite{GRETINA,Anderson:2009} and read-out using the ORCA data acquisition software~\cite{Howe:2004}. Timestamps are synchronized across the data acquisition system and signals from multiple detectors that occur within a 4~$\mu$s window are grouped. Events coincident with muons that trigger the external veto system are tagged for offline removal. Periods of high noise due to liquid-nitrogen fills are also excluded from the analysis. Throughout the data taking period, a bi-weekly, 4-hour energy calibration was performed with a \textsuperscript{228}Th line source~\cite{Abgrall:2017gpr}.

\begin{figure}[t]
 \centering
 \includegraphics[width=0.9\columnwidth]{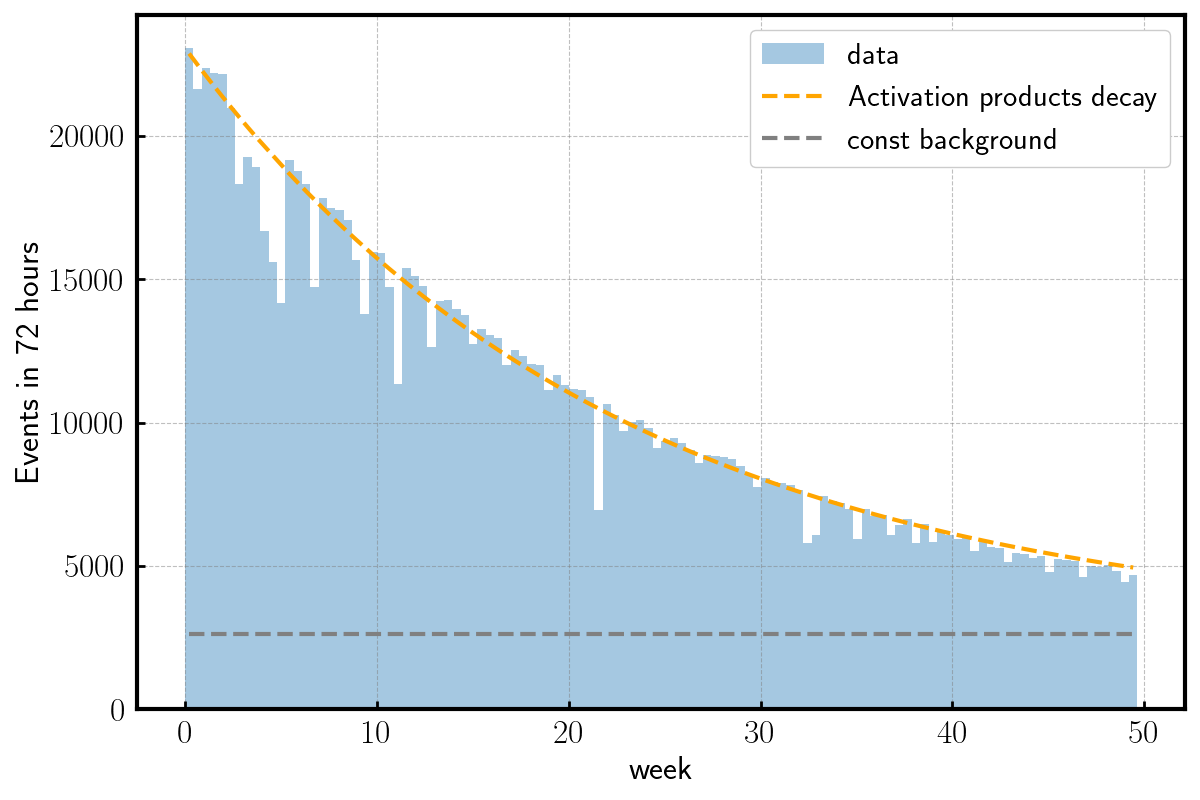}
 \caption{Count-rate in real time for the region between 100-500\,keV, where the \tam\ signatures are expected. The count rate, not life-time corrected, is due primarily to radioactivity in the Ta samples: \taH\ decay, the decay of other short-lived cosmogenic isotopes, and a constant rate from the U/Th decay chains. }
 \label{Fig:Rate}
\end{figure}

The \tam\ data analysis was done using the secondary analysis chain of \MJD, a Radware-based software package~\cite{Radware:ORNL}. Data from \textsuperscript{228}Th calibration data was used to set the energy scale for each subsequent two week period. The energy calibration procedure uses many of the tools developed for the \DEM, including pole-zero and charge-trapping corrections~\cite{Majorana:2022vai}. To estimate the quality of the energy calibration, the $\gamma$-rays from natural backgrounds (which are not used for calibration), including $^{182}$Ta, are fit with one or more Gaussian functions plus a linear background. For the lowest energies, an additional exponential background component is added to reproduce the rise of the spectrum towards lower energies. The fit is conducted over a $\pm$20\,keV window around the expected signal energy. For each energy, the width of the Gaussian agrees well with the resolution achieved during the neutrinoless double-beta decay search~\cite{Majorana:2022vai,Majorana:2019nbd,Majorana:2022udl}. Three of the twenty-three detectors showed gain drifts following a power outage that occurred midway through data taking, which negatively affected their energy resolution, and these detectors were not used in this analysis.

Following energy calibration, the data are checked for drop-out periods. This is done by measuring the event rate. If no events occur within a detector over a 4-hour window, that entire time period, for all detectors, is excluded from the analysis. Data collected during calibration runs were also removed from the \tam\ data-set, along with two, one-day shutdowns due to power outages at SURF. The array was live for 98.2\% of the data taking period as a result of these cuts.

 A 10-keV analysis threshold is applied to all data sets, and the data are blinded by removing events that fall within $\pm$2~keV of signature $\gamma$-rays. The possible transitions and the associated $\gamma$-rays energies are shown by the blue arrows in Fig.~\ref{fig:levels}: EC to $^{180}$Hf $\gamma$-rays are 93.3, 215.3, and 332.2\,keV; $\beta^{-}$ decay to $^{180}$W $\gamma$-rays are 103.6, 234.0, and 350.9\,keV; and internal $\gamma$-rays are 37.7 and 39.5\,keV~\cite{NucleDat180}. For the IC, only the 39.5\,keV transition can be observed. An additional signature of the $\gamma$ and IC branches is the observation of a 93.3 or a 103.6-keV $\gamma$-ray from the de-excitation of Hf or W, although the branching ratios to the first excited states of these nuclei is small (25\% for Hf and 4\% for W).

The total event rate of a few Hz observed in the detector array is dominated by signals originating from the Ta samples, see Fig.~\ref{Fig:Rate}. There is a constant event rate due to long-lived natural radioactivity in the discs and apparatus. From the \DEM\ data we can estimate that in the current configuration only about 10\% of the observed constant background comes from the latter, hence the sample disks contain around 0.5(1)~mBq \textsuperscript{238}U per kg Ta and 0.10(2)~mBq/kg \textsuperscript{232}Th. The decrease of the background rate is due to \taH\ and \hffive, which are the remnants of cosmogenic activation of the Ta samples above ground, with half-lives of 114 and 70.3 days, respectively. Previous studies stored their Ta samples underground for several years before beginning measurements to eliminate these backgrounds~\cite{Lehnert:2019tuw}.

\begin{figure}[h]
 \centering
 \includegraphics[width=1\columnwidth]{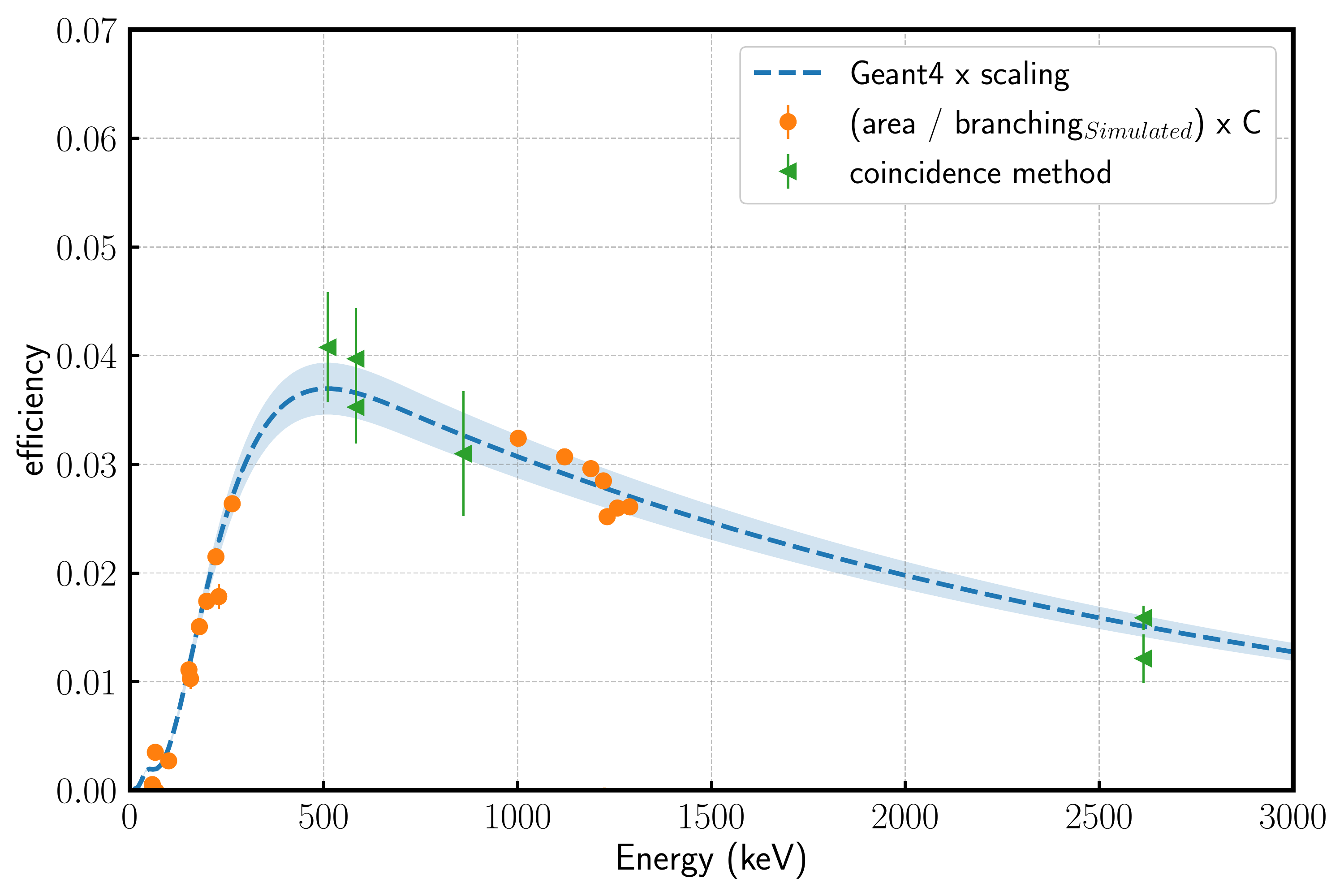}
 \caption{Simulated efficiency (blue dashed) compared to the intensities of $\gamma$-rays from \taH\ decays with branching ratio greater than 1\% (orange points). Since the absolute \taH\ activity is not known, the points are multiplied with an arbitrary constant $C$ to compare the distribution to the curve shape. The green points show efficiencies determined by the $^{208}$Tl coincidence method. The Geant4-derived curve, is normalized to these points (scaling factor 0.95(6)), whereas the band represents the uncertainty due to scaling.}
 \label{Fig:Eff}
\end{figure}

A crucial component of the half-life calculation is the efficiency for detecting the signature $\gamma$-rays emitted during the \tam\ decay. To determine this, a combination of experimental data and Monte Carlo simulation is used. First, a Geant4 simulation was performed in which individual $\gamma$-rays were emitted from uniformly distributed points within the Ta discs. The starting energy of the $\gamma$-rays was varied from 10~keV to 3~MeV in increments of 10~keV, and the efficiency of detecting these $\gamma$-rays in one of the detectors was calculated at each energy. The resulting interpolated efficiency curve is shown in Fig.~\ref{Fig:Eff}. The simulation assumes the cosmogenic activity is uniform between all of the Ta discs, which is consistent with the observed count rates in each detector.

The shape of the simulated efficiency curve is validated by a comparison with the observed intensities of the $\gamma$-rays from \taH\ decay, after correcting the signal intensities for the known branching ratios (e.g.~\cite{NucleDat182},) and including possible summing of multiple signals in one detector due to the close geometry. The absolute efficiency, or a possible scaling of the predicted curve, is determined using the coincidence method~\cite{Golovko:2022mgy,Hlavac:1999}. In this method, one compares the individual intensity of $\gamma$-rays in a cascade with the rate of multi-detector events to obtain the absolute efficiency of an individual detector. The $^{208}$Tl decay at the end of the natural $^{232}$Th-chain provides cascades that can be used for this analysis. Due to the low rate of $^{208}$Tl decays in the Ta discs, this method suffers from low statistics, especially for high multiplicity events, and the uncertainty on the derived efficiencies are large. The Geant4-derived efficiency is normalized to these points using a least-squares fit that results in a scaling of 0.95(6), and the efficiency values from this scaled curve are used in the following analysis.

The \tam\ half-life can be calculated from the following formula:
\begin{equation}
T_{1/2} = \ln 2 \frac{\epsilon_k}{S_k} N_{\text{Ta}} T_{\text{live}}
\label{Eq:2}
\end{equation}
where $S_k$ represents the counts in the $k^{th}$ decay channel, $\epsilon_k$ is the detection efficiency at the energy $E$ for specfic decay mode (shown by the curve in Fig.~\ref{Fig:Eff}), and $N_{\text{Ta}}$ is the number of \tam\ atoms, $6.84(17)\times 10^{21}$. The live time of the data-taking period, $T_{\text{live}}$, is 341.5 days.

\begin{figure}
 \centering
 \includegraphics[width=0.8\columnwidth]{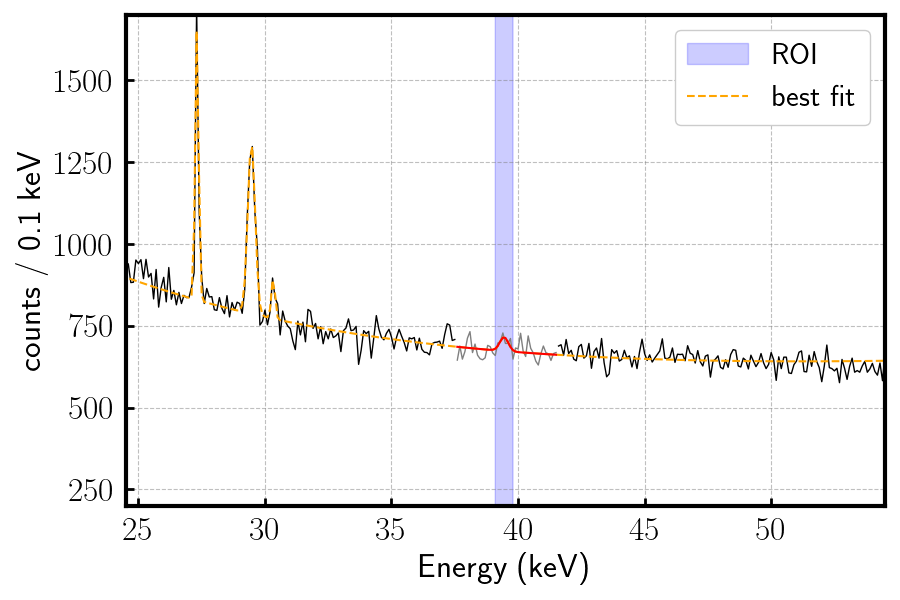}
 \includegraphics[width=0.8\columnwidth]{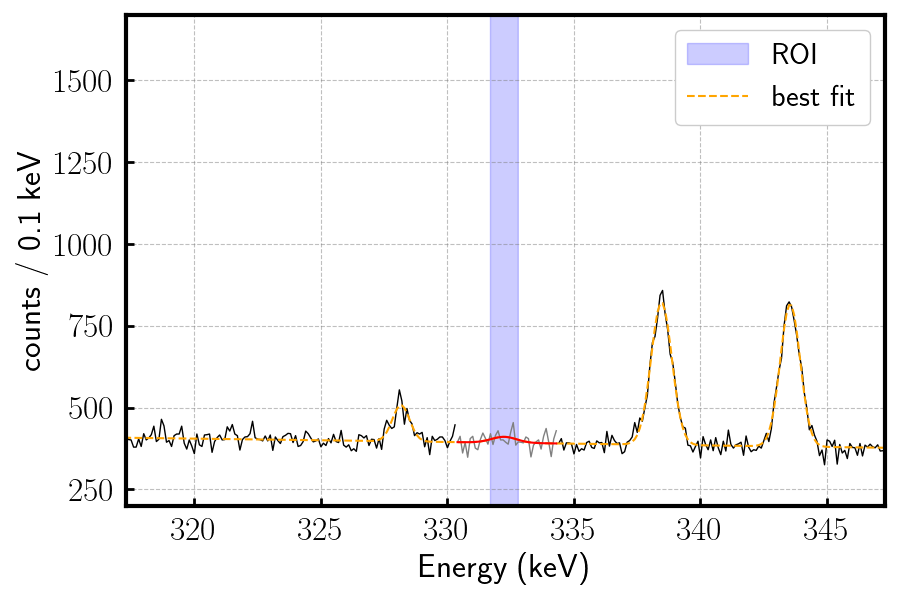}
 \caption{Regions of interest (ROI) for the 39.5-keV (top) and the 332.3-keV (bottom) $\gamma$-rays, respectively. The yellow line shows the best fit of the background peaks and flat background. The red curve shows the best fit of the signal peak.}
 \label{Fig:peak}
\end{figure}

A likelihood fit is used to extract the \tam\ decay signal strengths from the data. Spectral fits were performed in the regions of interest surrounding each of the characteristic $\gamma$-ray energies. The fits include a Gaussian peak shape for the signal, a linear background, and additional Gaussians at the energies of any known background lines in the region of interest, see e.g. Fig.~\ref{Fig:peak}. The literature value for the energy of the $\gamma$-rays and the expected energy resolution are used as initial values in the fit. The energy is allowed to float within $\pm0.5$~keV and the resolution is allowed to float within $\pm10$\% from the expected value. The background rate is fit to be about 0.7\,(0.5)~cts/keV/day averaged over the data-taking period in the 100-keV\,(300-keV) region and is comparable with previous experiments~\cite{Lehnert:2016iku}. The fit of the 93.3-keV and the 350.9-keV \tam\ signals are impacted by nearby background. The excellent energy resolution of the \DEM\ allows a simultaneous fit of multiple contributions from signal and backgrounds at known energies. Hence, all regions can be used but some will have larger uncertainties. Within all of the signal regions of interest, the best fit signal strength is within 2-$\sigma$ of a null-result. To calculate $S_k$, the best fit peak area plus 1.65-$\sigma$ (90\% C.L.) is used to calculate a limit on the decay rate.

In contrast to previous studies, the large number of detectors in close geometry combined with low background rate means a multiplicity analysis can be done that looks for the coincident $\gamma$-rays expected from some of the \tam\ decay channels. This analysis is competitive with fitting the single detector spectra because the reduction in signal detection efficiency (0.001 - 0.01) is counter-balanced by the improved background suppression ($\sim10^{-3}$ cts/keV/keV/day), so that the $\epsilon_k / S_k$ factor is similar to, or higher than, the simple spectral search. A two-dimensional histogram is made for events that contain two coincident energy deposits within the considered signal regions, and a two-dimensional likelihood fit is done assuming the same mean energy and peak resolution as in the one-dimensional fit. The efficiency $\epsilon$ from Eq.~\ref{Eq:2} now consists of the simulated detection efficiency for a two-detector event with the corresponding energies from within the cascade, cf. Fig.~\ref{fig:levels}. In this simulation two $\gamma$-rays are started with an angular correlation factor based on multipole momentum and spin of the emitting states~\cite{BohrMottelson}. The results from the spectral and two-dimensional fits are shown in Table~\ref{Tab:180Tanumbers}. These results improve upon the best existing limits for each decay channel and combine for a total half-life limit of $T_{1/2} > 0.67 \cdot 10^{18}$ years. In previous measurements~\cite{Hult:2006tp,Hult09,Lehnert:2016iku}, the total half-life is calculated without considering the isomeric transitions. Ref.~\cite{Hult09} does search for the isomeric transitions but does not include them in the total half-life calculation. The most recent work \cite{2305.17238} includes them, hence, the \MJD\ result represents an improvement of two orders of magnitude. For the direct decays of the isomeric state, the improvement in efficiency and reduction in background rate due to the coincidence method results in an improvement of about one orders of magnitude. These improvements are of great interest to the predictions on the basis of the K-selection rules~\cite{Ejiri:2017dro}.

\begin{table*}[t]
\footnotesize
\centering
\begin{tabular}{c | c c | c c | c c | c c | c c }
 & \multicolumn{2}{c}{${EC}$} & \multicolumn{2}{|c}{${\beta^-}$} & \multicolumn{2}{|c}{${\gamma}$} & \multicolumn{2}{|c}{${IC}$} & \multicolumn{2}{|c}{${\alpha}$} \\
 \hline
method & energy & $T_{1/2}$ & energy & $T_{1/2}$ & energy & $T_{1/2}$ & energy & $T_{1/2}$ & energy & $T_{1/2}$\\
 & (keV) & ($10^{18}$ yrs)) & (keV) & ($10^{18}$ yrs)) & (keV) & ($10^{18}$ yrs)) & (keV) & ($10^{18}$ yrs)) & (keV) & ($10^{18}$ yrs))\\
 \hline
SF & 93.3 & 1.23(30) & 103.6 & 1.54(17) & 37.7 & 0.63(8) & - & - & 184.1 & 4.80(42) \\
& 215.3 & 5.69(55) & 234.0 & 5.76(75) & 39.5 & 0.67(10) & 39.5 & 0.67(10) & 204.7 & 5.58(54) \\
& 332.2 & 10.0(13) & 350.9 & 9.31(114) & 93.3 & 0.29(4) & 93.3 & 0.29(4) & 388.8 & 10.2(12) \\
& & & & & 103.6 & 0.07(2) & 103.6 & 0.07(2) \\
\hline
2D & 93.3+215.3 & 1.88(35) & 103.6+234.0 & 2.65(49) & - & - & - & - & 184.1+204.7 & 11.3(22)\\
& 93.3+332.2 & 3.18(56) & 103.6+350.9 & 4.18(78) & - & - & - & - \\
& 215.3+332.2 & 13.3(22) & 234.0+350.9 & 15.4(27) & - & - & - & - \\
\hline
\hline
\textbf{Best - this work} & & 13.3(22) & & 15.4(27) & & 0.67(10) & & 0.67(10)& & 11.3(22) \\
previous works & & 1.6~\cite{2305.17238} & & 1.1~\cite{2305.17238} & & 0.0045~\cite{2305.17238} & & 0.0045~\cite{2305.17238} & & - \\
\hline
Expected $T_{1/2}$ \cite{Auerbach:2017, Ejiri:2017dro, ROYER2010279, Fenyes1964} & & 10$^{23}$yrs & & 10$^{20}$yrs & & 10$^{31}$yrs & & 10$^{18-19}$yrs & & 10$^{28}$yrs\\
\end{tabular}
\caption{Measured decay half-life limits. Results are given at a 90\% C.L. using the one-dimensional spectral fits (SF), a multiplicity-two analysis (2D) where applicable, and the strongest limit for the decay channel. The nomenclature introduced in Eq.~\ref{Eq:1} is used to describe each decay channel.}
\label{Tab:180Tanumbers}
\end{table*}

\begin{figure}[t!]
 \centering
 \includegraphics[width=1\columnwidth]{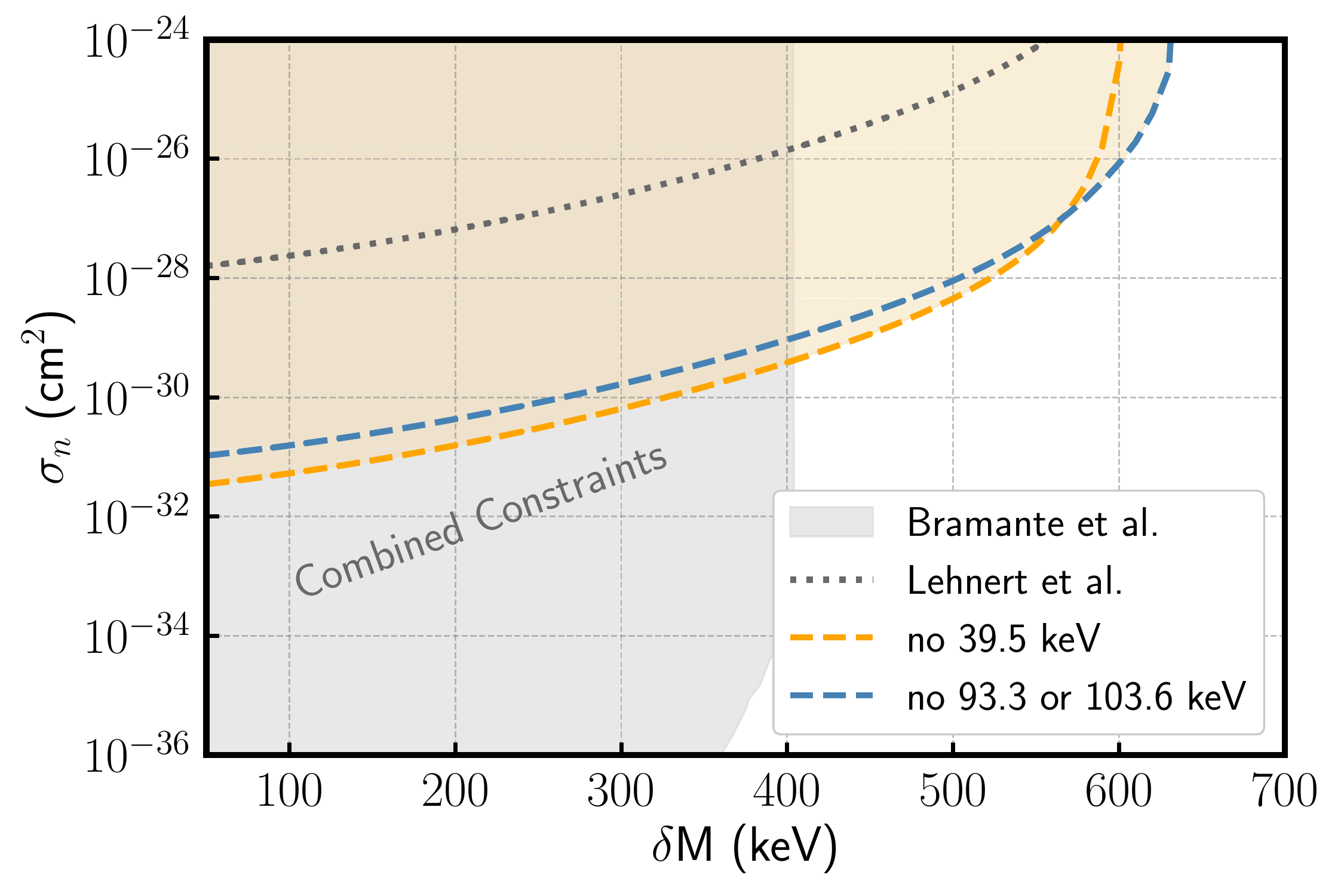}
 \includegraphics[width=1\columnwidth]{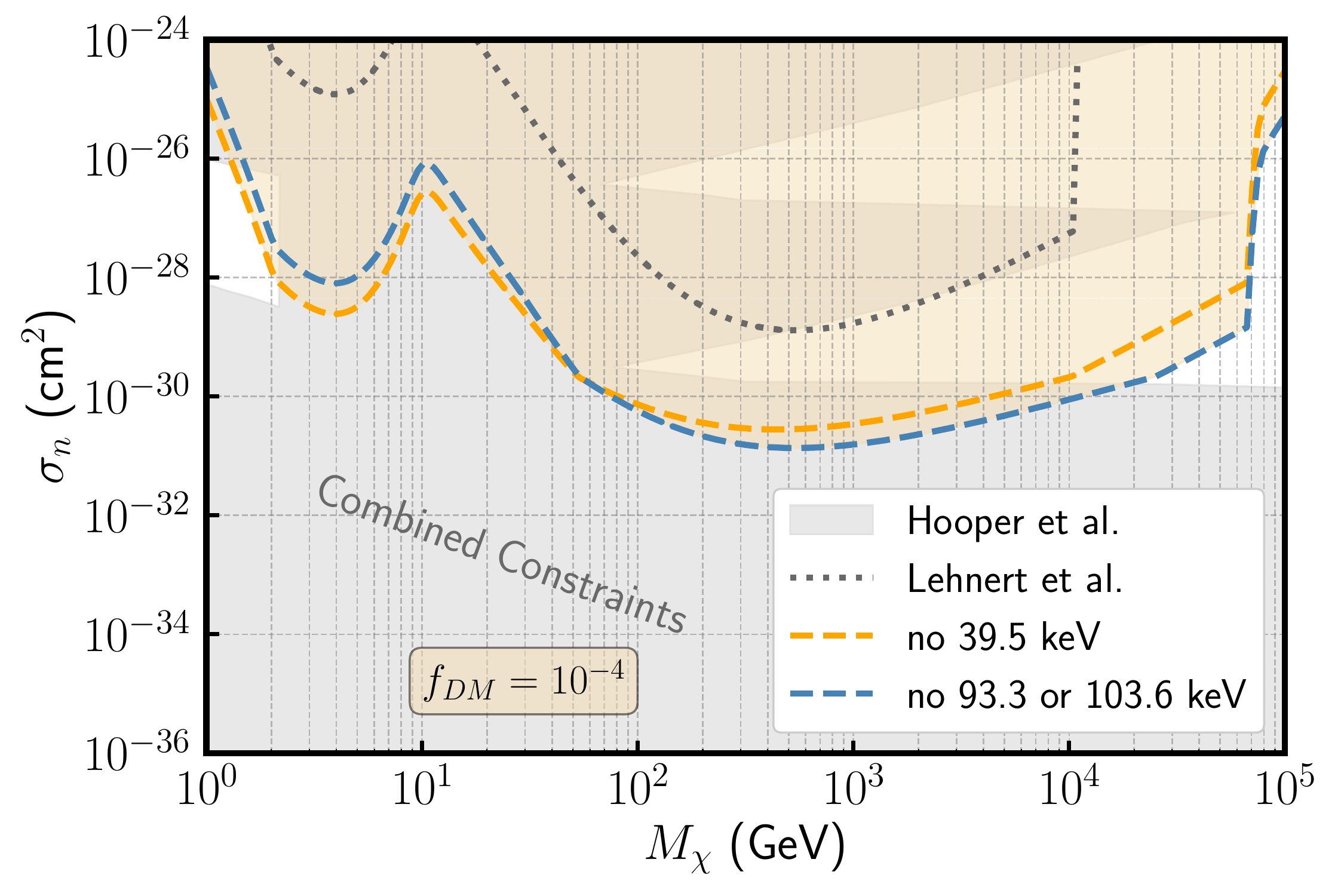}
\caption{(top) 90\% C.L. exclusion limits on the per-nucleon cross-section for strongly interacting DM assuming a fractional relic density of $f_{DM}=10^{-4}$. The limit based on the non-observation of the 93.3~keV line (blue dashed) is sensitive to DM-induced de-excitations to the \tag\ first excited ($2^+$) and ground ($1^+$) states. This can be compared to limits from Ref.~\cite{Lehnert:2019tuw} (grey dashed), which is based on the 103.5~keV signature. The limit derived from the non-observation of the 39.5~keV line (dashed orange) is only sensitive to DM-induced de-excitations to the first excited state. Both cover phase space not covered by other experimental approaches \cite{Hooper}. (bottom) 90\% C.L. exclusion limits on the per-nucleon cross-section for inelastic DM with mass splitting $\delta M$. Color coding is identical to the top plot (gray shaded region is based on Ref.~\cite{Bramante}).}
 \label{Fig:DM}
\end{figure}

The non-observation of the transition to the ground state decays ($\Gamma_\gamma$ and $\Gamma_{IC}$) constrains the phase-space of certain classes of DM models that evade traditional underground detection methods~\cite{Pospelov2020,Lehnert:2019tuw}, cf. Fig~\ref{Fig:DM}. Strongly interacting DM, which thermalizes as it passes through the earth, rendering it undetectable via nuclear scattering, would mediate exothermic transitions from the \tam\ state and measurably increase the decay rate of the isomer. Similarly, in inelastic DM models, where ground state DM only interacts inelastically with the Standard Model, and upscatters to an excited state by downscattering \tam\ increasing the measured \tam\ decay rate.

With 341.5 live days of data, we set new limits on the the decay of \tam\ decay that improve upon previous half-lifes by two to three orders of magnitude. We also derive new constraints on strongly interacting and inelastic dark matter. Data-taking with the \DEM\ array will continue into 2024, and as the background rate decreases further to about a quarter of the current average due to the decay of cosmogenics within the Ta samples, sensitivity will continue to improve to levels that are comparable to Ref.~\cite{2305.17238}. Besides dedicated $\beta\beta$-searches and some $\alpha$ decays, the results presented are the most sensitive search for radioactive decays ever achieved.

\begin{acknowledgements}
We gratefully acknowledge that the research presented in this report was supported by the Laboratory Directed Research and Development program of Los Alamos National Laboratory under project number 20220092ER, which enabled the \tam\ rare decay search.
This material is based upon work supported by the U.S.~Department of Energy, Office of Science, Office of Nuclear Physics under contract / award numbers DE-AC02-05CH11231, DE-AC05-00OR22725, DE-AC05-76RL0130, DE-FG02-97ER41020, DE-FG02-97ER41033, DE-FG02-97ER41041, DE-SC0012612, DE-SC0014445, DE-SC0018060, DE-SC0022339, and LANLEM77/LANLEM78. We acknowledge support from the Particle Astrophysics Program and Nuclear Physics Program of the National Science Foundation through grant numbers MRI-0923142, PHY-1003399, PHY-1102292, PHY-1206314, PHY-1614611, PHY-1812409, PHY-1812356, PHY-2111140, and PHY-2209530. We gratefully acknowledge the support of the Laboratory Directed Research \& Development (LDRD) program at Lawrence Berkeley National Laboratory for this work. We gratefully acknowledge the support of the U.S.~Department of Energy through the Los Alamos National Laboratory LDRD Program, the Oak Ridge National Laboratory LDRD Program, and the Pacific Northwest National Laboratory LDRD Program for this work. We gratefully acknowledge the support of the South Dakota Board of Regents Competitive Research Grant.
We acknowledge the support of the Natural Sciences and Engineering Research Council of Canada, funding reference number SAPIN-2017-00023, and from the Canada Foundation for Innovation John R.~Evans Leaders Fund.
We acknowledge support from the 2020/2021 L'Or\'eal-UNESCO for Women in Science Programme.
This research used resources provided by the Oak Ridge Leadership Computing Facility at Oak Ridge National Laboratory and by the National Energy Research Scientific Computing Center, a U.S.~Department of Energy Office of Science User Facility. We thank our hosts and colleagues at the Sanford Underground Research Facility for their support.

\end{acknowledgements}

\bibliography{Tantalum.bib}

\end{document}